\documentclass[showpacs,preprintnumbers,amsmath,amssymb]{revtex4}
\usepackage{graphicx}% Include figure files
\usepackage{dcolumn}% Align table columns on decimal point
\usepackage{bm}% bold math

\begin{document}
\draft

\preprint{Submitted to Phys.~Rev.~Lett.(January 2007)}

\title{Guiding, focusing, and sensing on the sub-wavelength scale
using metallic wire arrays}

\author{G.~Shvets}
\email{gena@physics.utexas.edu}
\author{S.~Trendafilov}
\affiliation{Department of Physics, The University of Texas at
Austin, Austin, Texas 78712 }
\author{J.~B.~Pendry}
\affiliation{Blackett Laboratory, Imperial College, Prince Consort
Road, London SW7 2BW, United Kingdom}
\author{A.~Sarychev}
\affiliation{Ethertronics Inc., San Diego, CA 92121}

\date{\today}% It is always \today, today,

\begin{abstract}
We show that two-dimensional arrays of thin metallic wires can
guide transverse electromagnetic (TEM) waves and focus them to the
spatial dimensions much smaller that the vacuum wavelength. This
guiding property is retained for the tapered wire bundles which
can be used as multi-channel TEM endoscopes: they capture a
detailed electromagnetic field profile created by deeply
sub-wavelength features of the studied sample and magnify it for
observation. The resulting imaging method is superior to the
conventional scanning microscopy because of the parallel nature of
the image acquisition by multiple metal wires. Possible
applications include terahertz and mid-infrared endoscopy with
nanoscale resolution.

\end{abstract}

\pacs{}% PACS, the Physics and Astronomy
\keywords{Surface plasmons, sub-wavelength optics}%Use showkeys class option if keyword
\maketitle

Diffraction of light is the major obstacle to a variety of
applications requiring concentrating optical energy in a small
spatial volume: light cannot be confined to dimensions much
smaller than half of its wavelength $\lambda/2$. Applications that
would benefit from overcoming the diffraction limit include
nonlinear spectroscopy and harmonics
generation~\cite{feld_prl97,brus_jpcb00,ichimura_prl04,xie_opl01},
sub-wavelength optical
waveguiding~\cite{berini_prb01,bozhevolniy_nature06,atwater_slot_nano06},
and nanofabrication~\cite{shaochen_apl05}. Utilizing plasmonic
materials with a negative dielectric permittivity circumvents
diffraction limit because interfaces between polaritonic
($\epsilon < 0$) and dielectric ($\epsilon > 0$) materials support
surface plasmons that can be confined to sub-$\lambda$ dimensions.
Examples of diffraction-beating devices based on plasmonics
include
superlenses~\cite{pendrylens_prl00,zhang_sci05,melville_optexp05,taubner_science06},
coupled-sphere waveguides~\cite{atwater_nanosphere_apl01}, and
sharp focusing tips~\cite{stockman_prl04}.

\begin{figure}
\includegraphics[width=35mm]{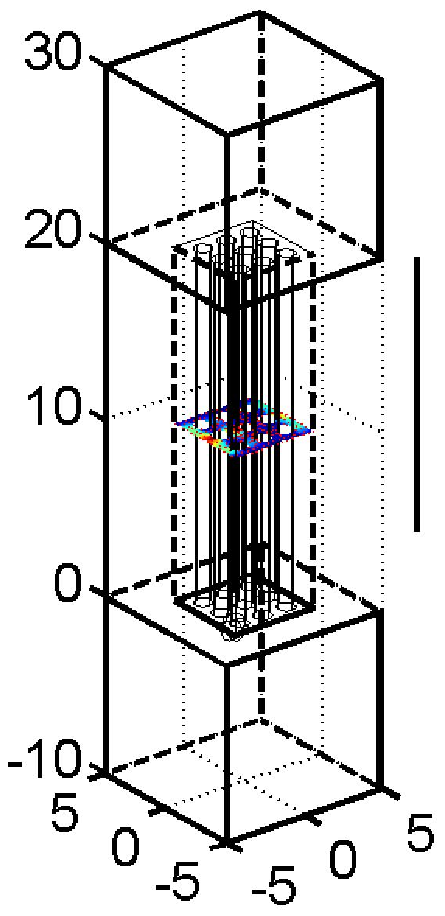}

\vspace{170pt}

\includegraphics[width=35mm]{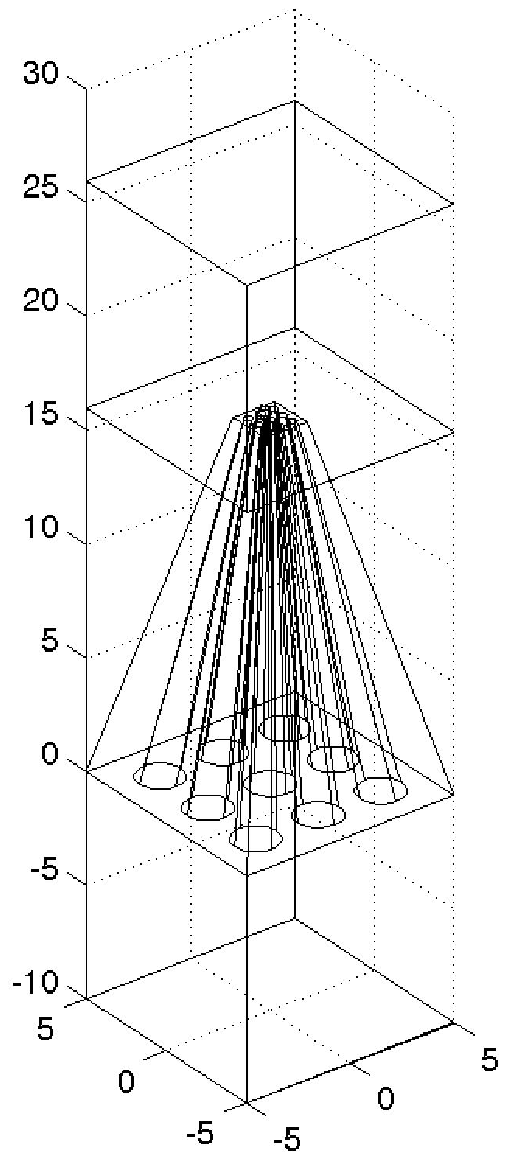}
 \caption{\label{fig:endoscope_schematic} Schematics of two
 sub-wavelength endoscopes based on a $3 \times 3$ array of metal
 wires embedded in a straight (left) or tapered (right) metal-coated
 fiber. Both endoscopes are terminated by square single-mode
 rectangular waveguides on both ends.}
\end{figure}

\begin{figure}
\includegraphics[width=65mm]{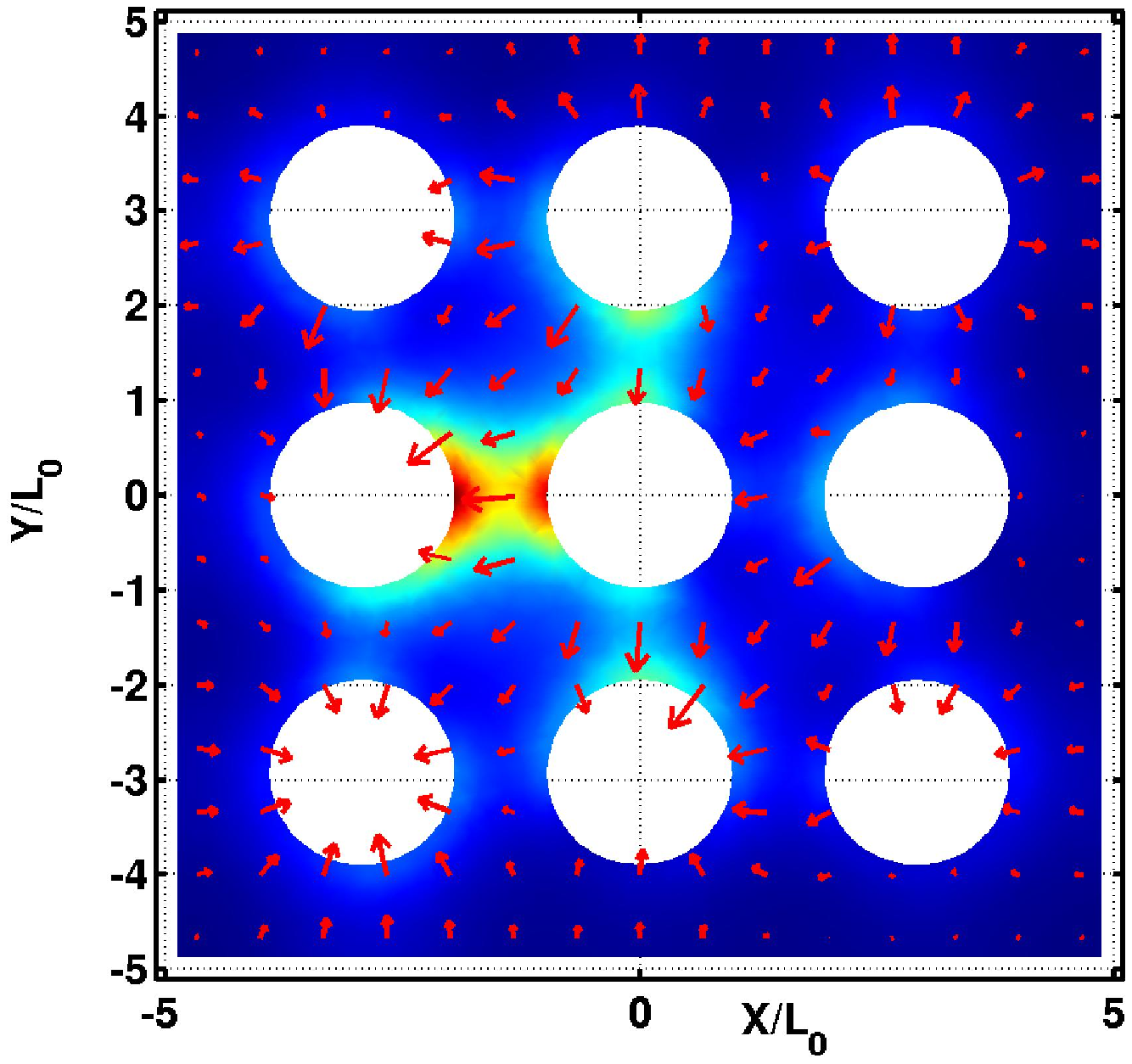} \hspace{40pt}
\includegraphics[width=65mm]{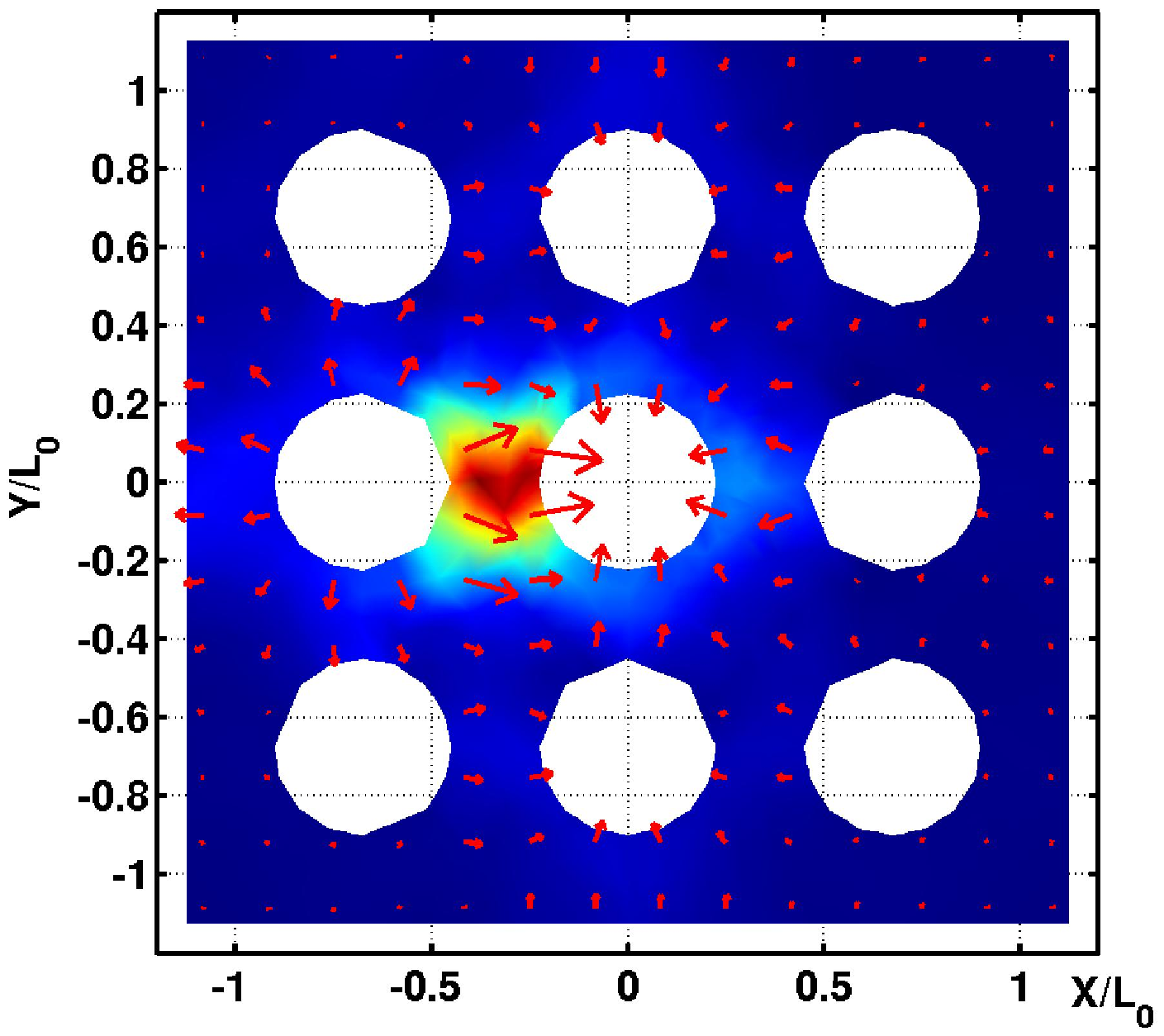}
 \caption{\label{fig:cone_magn_demagn} (Color online) Applications
 of a tapered endoscope from
 Fig.~\ref{fig:endoscope_schematic}(right): image magnification
 and de-magnification by a factor $5$. (a) Image magnification:
 image of a small metal sphere (diameter $D_{\rm small} = \lambda/25$) placed
 just above the tip at $(x_{obj} = -D_{\rm small}/2, y_{obj} = 0)$ is transported to the base
 plane. (b) Image de-magnification: image of a larger metal sphere (diameter
 $D_{\rm large} = \lambda/5$) placed just below the base at $(x_{obj} = D_{\rm large}/2, y_{obj} =
 0)$ is transported to the tip.}
\end{figure}

High losses associated with surface plasmonics are hampering many
of these applications. Another challenge yet to be met is
designing practical imaging modalities based on sub-$\lambda$
plasmons that convert near-field electromagnetic (EM)
perturbations into the far field where they can be easily
observed. In this Letter we propose a solution to these two
problems: a tapered multi-wire array supporting sub-wavelength
transverse electromagnetic (TEM) waves. Examples of the multi-wire
endoscopes based on such arrays (un-tapered and tapered) are shown
in Fig.~\ref{fig:endoscope_schematic}. We have demonstrated that
the tapered endoscope can accomplish two tasks: (i) creating near
the {\it base} of an endoscope a magnified image of deeply
sub-wavelength objects (metal spheres, in our case) placed at the
endoscope's {\it tip} [see Fig.~\ref{fig:cone_magn_demagn}(a)],
and (ii) creating near the {\it tip} of an endoscope a
de-magnified image of a mask placed at the endoscope's {\it base}
[see Fig.~\ref{fig:cone_magn_demagn}(b)]. Accomplishing the first
task is necessary for making a sub-$\lambda$ sensor while
accomplishing the second one -- for making a sub-$\lambda$
lithographic tool.

%Various spectroscopic applications also benefit from the broad
%bandwidth of TEM waves. We show in this Letter we present a new

%Because TEM waves are not relying on any resonances, they can have
%a broad bandwidth: yet another benefit for spectroscopic
%applications.
%problem of conversion between near and far fields which does not
%suffer from the high losses associated with surface plasmons:

Single metallic wires and coaxial wire cones have recently
attracted considerable attention as low-loss
waveguides~\cite{keilmann_95,mittleman_nature04} of TEM-like modes
of THz and far-infrared radiation. Using a single wire waveguide
has its limitations: for example, if a wire is used as a high
spatial resolution sensor, then only a single bit of information
can be collected without scanning the wire. We demonstrate that a
bundle of closely spaced wires can act as a multi-channel sensor
capable of simultaneously collecting information from a spatially
distributed object. Electromagnetic properties of metallic wire
arrays has been previously investigated in the context of
metamaterials~\cite{pendry_plasmon96,sarychev_wires00,belov_wires03,shapiro_optlett06}.
Below we review the electromagnetic properties of an infinite
square array with period $d$ of straight (along the $z$-direction)
metallic wires of an arbitrary shape in the $x-y$ plane. Two
well-known types of EM waves with the propagation wavenumber $k_z$
and frequency $\omega$ characterized by the scalar potentials
$\phi$ and $\psi$ are supported: (i) transverse magnetic (TM)
modes with $\vec{B} = \vec{e}_z \times \vec{\nabla} \phi$, $E_z =
i c \nabla_{\perp}^2 \phi /\omega,$ $\vec{E}_{\perp} = c k_z
\vec{\nabla}_{\perp} \phi /\omega$, and (ii) transverse electric
(TE) modes with $\vec{E} = \vec{e}_z \times \vec{\nabla} \psi$,
$B_z = -i c \nabla_{\perp}^2 \psi /\omega$, and $\vec{B}_{\perp} =
-\c k_z\vec{\nabla}_{\perp} \psi /\omega$. Here
$\psi(\vec{x}_{\perp})$ and $\phi(\vec{x}_{\perp})$ satisfy the
following differential equation: $\left( \nabla_{\perp}^2 +
\omega^2/c^2 - k_z^2 \right) \{\psi,\phi \} = 0$. In addition to
$k_z$, these waves are characterized by the transverse Bloch
wavenumber $\vec{k}_{\perp} \equiv k_x \vec{e}_x + k_y \vec{e}_y$,
where $\Phi = \{ \phi,\psi \}$ satisfies the phase-shifted
boundary conditions: $\Phi(x=d/2,y) = \exp{(ik_x d)}
\Phi(x=-d/2,y)$ and $\Phi(x,y=d/2) = \exp{(ik_y d)}
\Phi(x,y=-d/2)$. At the perfect metal surface $\phi=0$ and
$\psi=const$ are also satisfied. Dispersion relations
$\omega_{TE,TM}^{(n)}(k_z,\vec{k}_{\perp})$, where $n$ labels the
propagation band (the Brillouin zone) can be readily computed
numerically for both types of waves.

The lesser known wave type is the TEM eigenmode of the periodic
metal cylinder array with a very simple dispersion relation:
$\omega^2(k_z,\vec{k}_{\perp}) \equiv k_z^2 c^2$. (If the medium
between the wires has a dielectric permittivity $\epsilon_d$, then
$\epsilon_d \omega^2 \equiv k_z^2 c^2$). TEM waves have no
longitudinal electric or magnetic fields, and are characterized by
a single scalar potential
$\phi(\vec{x}_{\perp})_{\vec{k}_{\perp}}$ satisfying the
phase-shifted periodic boundary conditions and $\phi = const.$ at
the metal surface. Electric and magnetic field of the TEM wave are
orthogonal to each other and given by
$\vec{E}=\vec{\nabla}_{\perp} \phi$ and $\vec{B} = \vec{e}_z
\times \vec{\nabla}_{\perp} \phi$. As will be shown below, the
remarkable property of the TEM waves of being dispersionless with
respect to the transverse wavenumber $\vec{k}_{\perp}$ can be
explored in sub-wavelength guiding/imaging applications. One can,
therefore, view TEM modes as being {\it transversely local}: the
image of an object with finite transverse size does not spread out
as it is transported along the endoscope. To understand why the
propagation wavenumber of TEM modes is degenerate in
$\vec{k}_{\perp}$, consider a {\it finite} $N \times N$ array of
wires surrounded by a perfectly conducting metal shell. Finite
extent of the array discretizes $\vec{k}_{\perp}$ so that there
are $N^2$ distinct wavenumbers supported by the array. This is in
agreement with the well known fact~\cite{carbonini_ieee92} that
the vector space of TEM potentials of a multiconnected coaxial
waveguide has the dimension equal to the number of inner
conductors. Therefore, TEM modes of an endoscope consisting an $N
\times N$ wire bundle are capable of transferring $N^2$ channels
of information along its length.

An ideal endoscope transfers an arbitrary image of the field
distribution at $z=0$ over a significant distance to $z=L$ with
minimal distortion. Indeed, any field distribution at $z=0$ with
an arbitrary spatial detail size $\Delta << d$ can be expanded as
the sum of TE, TM, and TEM modes with the Bloch wavenumbers
$\vec{k}_{\perp}$ and the Brillouin zone index $n$, and propagated
with their corresponding propagation constants $k_{z}^{(i,n)}$,
where $i=1,2,3$ correspond to TEM, TE, and TM modes, respectively.
However, if the total crossection of the tapered endoscope becomes
smaller than $\lambda^2/4$, then all TE and TM modes are
evanescent, i.~e.~$k_{z}^{2(i,n)} < 0$ for $i=2,3$. The only modes
that can transport the image without distortion are the TEM modes.
Because they do not form a complete set, they can only ensure
spatial resolution of order the wire spacing $d$. Therefore,
imaging with TEM waves is a form of discrete sampling: the exact
spatial profile of a small scatterer with a spatial dimension
$\Delta << d$ will not be resolved in the image, but its presence
in a specific $d \times d$ unit cell will be detected. Because TEM
modes have no cutoff, making the spacing $d$ extremely
sub-wavelength results in an arbitrary high spatial resolution.

%TE modes become cut off for $\omega < \omega_p$, where $\omega_p =
%c/d \left[ 2\pi/\ln{d/w} \right]$ is the effective "plasma
%frequency"~\cite{pendry_plasmon96}, and $w$ is the wire diameter.
%Even if the total crossection of the wire bundle is not
%sub-wavelength, the TM modes that are not cut off are going to be
%highly dispersive with respect to $\vec{k}_{\perp}$. Therefore,
%the image transferred by TM modes is going to be highly distorted,
%and will contain a very limited amount of information due to the
%small number of propagating modes.

\begin{figure}
\includegraphics[width=65mm]{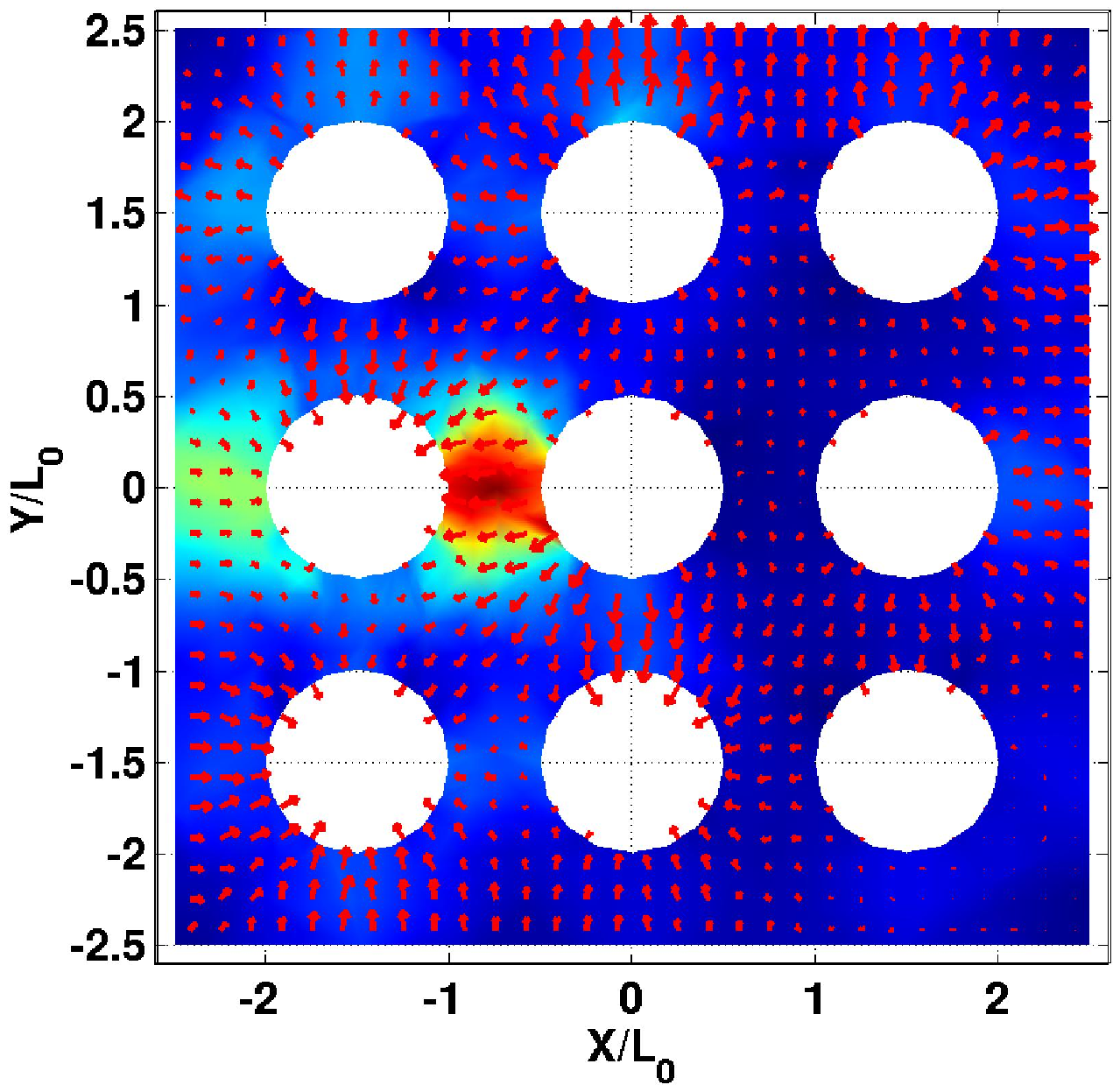} \hspace{70pt}
\includegraphics[width=65mm]{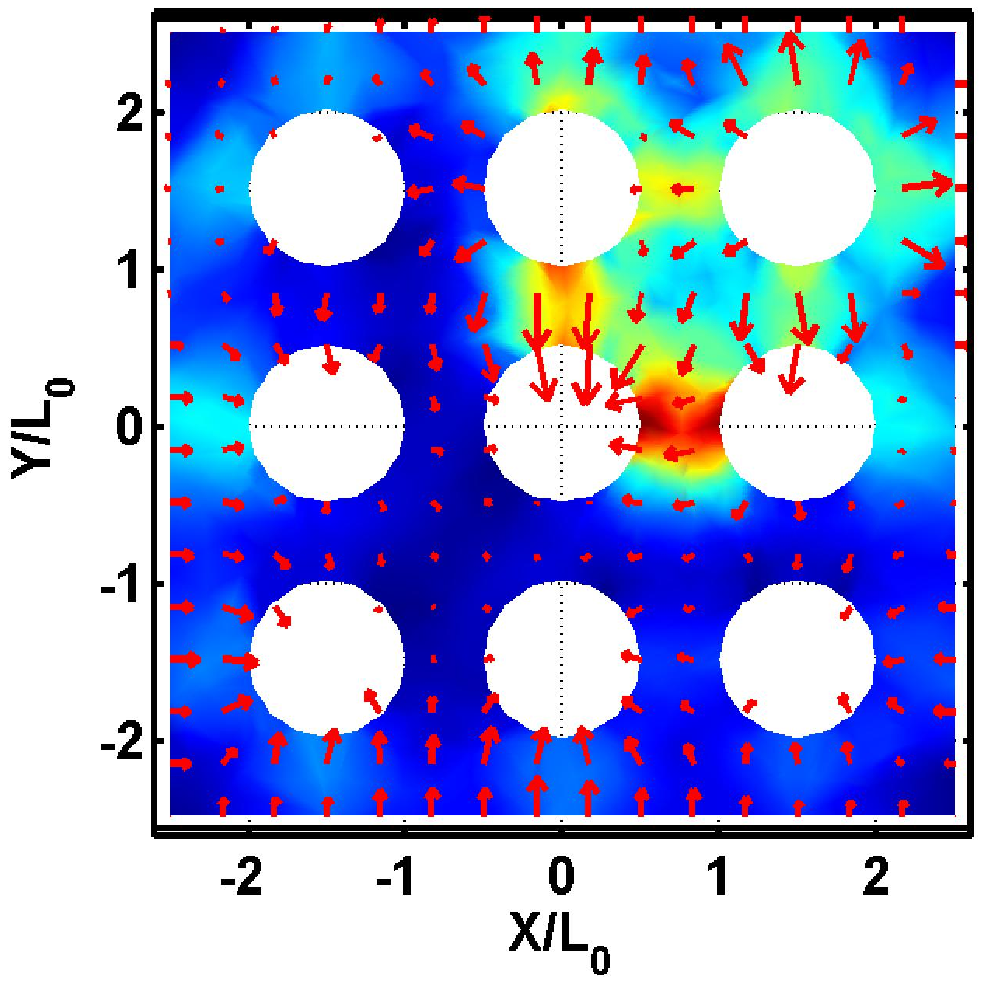}
 \caption{\label{fig:endo_nocone} (Color online) Transport of the
 image of a metal sphere (diameter $D = \lambda/10$) by a straight sub-wavelength
 endoscope shown in Fig.~\ref{fig:endoscope_schematic}(left) consisting of a $3
 \times 3$ wire array placed inside a square sub-wavelength metallic waveguide of
 the width $W = \lambda/3$. The sphere's center is at $z_{obj} = -0.7D$,
 $x=x_{obj}$, $y=y_{obj}$. Shown are the color-coded $|\vec{E}_{\perp}|^2$
 profiles in the imaging plane $z=4\lambda/3$ for (a) $(x_{obj} = -d/2, y_{obj} =
 0)$, and (b) $(x_{obj} = d/2, y_{obj} = d/2)$. Arrows represent the electric
 field.}
\end{figure}

To demonstrate how a metal wire endoscope can transport a deeply
sub-wavelength image, we have numerically simulated the following
problem: transferring an image of a metallic sphere with a
diameter $D = \lambda/10$ using a $3 \times 3$ array of conducting
wires encased in a square $\lambda/3 \times \lambda/3$
sub-wavelength metal waveguide. Wire spacing and diameter are $d =
\lambda/10$ and $w=\lambda/15$, endoscope's length is $L =
4\lambda/3$. All simulations in this Letter are made under a
simplifying assumption of perfectly electrically conducting (PEC)
metals. As shown at the end of the Letter, this assumption is
valid for EM waves spanning mid-IR and THz frequency ranges. PEC
boundary conditions make the results scalable to any wavelength.
Therefore, all dimensions are scaled to an arbitrary length scale
$L_0 = \lambda/15$. Dielectric permittivity of the surrounding
medium was assumed to be $\epsilon_d = 1$. The schematic of the
endoscope is shown in Fig.~\ref{fig:endoscope_schematic}(left).
The EM wave is launched from a single-mode square $2\lambda/3
\times 2 \lambda/3$ waveguide at $z=-10L_0$. We have chosen a
circularly polarized incident wave to avoid polarization
sensitivity of a square array of wires. The scattering metal
sphere's center is at $z_{obj} = -0.7D$, $x=x_{obj}$, $y=y_{obj}$.
Two lateral sphere positions have been simulated: (a) $(x_{obj} =
-d/2, y_{obj} = 0)$, and (b) $(x_{obj} = d/2, y_{obj} = d/2)$. The
respective intensity distributions of the $|\vec{E}_{\perp}|^2$ at
the end of the endoscope ($z=19L_0$) shown in
Figs.~\ref{fig:endo_nocone}(a,b) confirm the earlier made
statement about the sampling nature of TEM-based imaging: only the
mere presence of a scattering sphere inside a given elementary
cell is detected, with the details of the scatterer's shape lost.
Nevertheless, the spatial resolution equal to the size of the unit
cell $d = \lambda/10$ is clearly established by this simulation.
The peak intensity in the imaging plane is higher by one order of
magnitude when the scattering object is present compared with the
case of a multi-wire endoscope with no scattering object:
$I_{scatt}/I_{wire} = 10$. The latter intensity is another five
orders of magnitude higher than when the wires are removed from
the waveguide: $I_{wire}/I_{wg} = 10^5$.

%For example, while Fig.~1(b) has a bright intensity spot exactly
%where the scattering sphere is (between the two wires), Fig.~1(c)
%has bright regions surrounding the scattering spheres' position (in
%the middle of the cell).

Next, we demonstrate that an endoscope based on a {\it tapered}
metal wire array shown in
Fig.~\ref{fig:endoscope_schematic}(right) is capable of
magnification and demagnification. One obvious application of
image magnification is a sensor collecting EM fields from highly
sub-wavelength objects in the immediate proximity of the
endoscope's tip and transforming them into a much larger
detectable image. Image demagnification can be applied to surface
pattering and lithography: a complex large mask can be placed
close to the wide base of the endoscope and projected/focused
towards the tip creating a highly sub-wavelength intensity
distribution in the tip's vicinity. We've simulated a
pyramid-shaped metallized fiber threaded by a $3 \times 3$ array
of metallic wires. Endoscope's base has a $10L_0 \times 10L_0$
square crossection (where, as before, $L_0 = \lambda/15$), wires
separation is $d=3L_0$, wires' diameters are $w=2L_0$. All these
dimensions are proportionately scaled down by a factor $5$ at the
tip. The purpose of this simulation is to illustrate image
magnification and demagnification by a factor $5$. As in the
non-tapered case, the tapered endoscope is terminated on both ends
by a single-mode ($2\lambda/3 \times 2\lambda/3$) metallic
waveguide. A practical multi-channel endoscope will have a much
larger (e.~g., $25 \times 25$) number of metal wires.

For magnification demonstration, a small metallic sphere with
diameter $D_{\rm small} = \lambda/25$ is placed at a distance
$\Delta z = 0.7D_{\rm small}$ above the endoscope's tip half-way
between the central wire and the next one on the left. The sphere
is illuminated from the top by a circularly polarized
electromagnetic wave. The image of $|\vec{E}_{\perp}|^2$ taken at
$z_{im} = L_0$ (slightly above the endoscope's base) is shown in
Fig.~\ref{fig:cone_magn_demagn}(a). The sphere's image (or that of
any strong scatterer) magnified by a factor $5$ appears as an
enhanced field in the image plane. The following intensity
contrasts are found: $I_{scatt}/I_{wires} = 3$ and
$I_{wires}/I_{wg} = 10^{3}$.
%Thus, we have demonstrated that
%a subwavelength object placed near the tip of a tapered endoscope
%can be magnified by a factor $5$ near the base of the endoscope.

The opposite process (de-magnification, or image focusing) can
also be demonstrated using the same tapered endoscope. A metallic
sphere with the diameter $D_{\rm large} = \lambda/5$ is placed at
a distance $\Delta z = 0.7D_{\rm large}$ below the endoscope's
base half-way between the central wire and the next one on the
left. The image located in the plane of the tip (hot spot shown in
Fig.~\ref{fig:cone_magn_demagn}(b)) is spatially compressed by a
factor $5$. Despite the fact that the electromagnetic wave
propagates through a very narrow waveguide, field intensity in the
hot spot is about the same as that of the incident wave. Had the
coupling efficiency of the incident wave into TEM waves been close
to unity, one would expect an intensity increase by a factor $25$
due to the narrowing of the endoscope's area. That this is not
happening is attributed to the low coupling efficiency because of
the sub-wavelength size of the scattering sphere. Nevertheless,
this simulation illustrates that extremely sub-wavelength
intensity landscapes can be created near the tip of a tapered
nanowire array. The following intensity contrasts are found:
$I_{scatt}/I_{wires} = 15$ and $I_{wires}/I_{wg} = 10^{5}$.

All simulations presented in this Letter were performed using the
PEC assumption. This assumption is highly accurate in the
far-infrared and THz frequency ranges. It is, however, instructive
to check whether the concept of a multi-wire endoscope could be
potentially extended to mid-infrared wavelengths. Below we
demonstrate that electromagnetic modes of an array of {\it
plasmonic} wires closely resembling TEM modes of an array of {\it
PEC} wires do exist. These surface plasmon polariton (SPP) modes
possess two essential properties enabling them to guide, focus,
and perform local sensing on a nanoscale: (a) they are low loss,
and (b) they are essentially dispersionless in the transsverse
direction, i.~e.~$\omega^2/c^2 \approx \alpha k_z^2 + \beta
\vec{k}_{\perp}^2$, where $\beta \ll \alpha$. Let's consider
$\lambda = 5 \mu m$ because of the importance of this wavelength
to chemical sensing~\cite{brehm_nanolett06} as well the
availability of low-loss silica fibers. We have used a commercial
finite elements code COMSOL to compute the propagation constants
of such SPPs assuming a square array of gold wires ($d = 0.5 \mu
m$, $w = 0.33 \mu m$, $\epsilon_{\rm Au} = -916 + 228i$) embedded
in a silica fiber with $\epsilon_d = 2.25$. An endoscope based on
this wire array provides $\lambda/10$ spatial resolution. For the
center of the Brillouin zone it was found that $c
k_z(\vec{k}_{\perp}=0)/\sqrt{\epsilon_d}\omega = 1.12 + 0.01i
\equiv \chi_r + i \chi_{im}$ confirming low loss of the TEM-like
SPPs. Very weak dispersion of $k_z$ on $\vec{k}_{\perp}$ of the
SPPs was confirmed by calculating $k_z(k_x=\pi/d)$ at the edge of
the Brillouin zone: $G \equiv [k_z(k_x=\pi/d) -
k_z(\vec{k}_{\perp}=0)]/k_z(\vec{k}_{\perp}=0) = 7 \times
10^{-3}$.

The validity of the ideal TEM description is justified for
transport distances of $L < \lambda/2\pi\epsilon_d \times
min(\pi/G,1/\chi_{im})$. Ohmic losses over distances $L >
\lambda/2\pi\epsilon_d/\chi_{im} \approx 8\lambda$ reduce the
transmitted light intensity but do not necessarily deteriorate the
spatial resolution ($\lambda/10$) of the image. Transverse
dispersion, however, reduces spatial resolution below $\lambda/10$
for $L > 42 \lambda$. For higher spatial resolutions, however,
transverse dispersion become more severe than Ohmic losses: an
endoscope must be shorter than $L = 5.5\lambda$ if the spatial
resolution of $\lambda/25$ is desired. We conclude from these
results that, although the classic dispersion relation $k_z =
\sqrt{\epsilon_d} \omega/c$ for TEM waves is no longer strictly
satisfied for plasmonic wires, the TEM-like SPPs are sufficiently
low-loss and dispersionless that the performance of the un-tapered
and tapered multi-wire endoscopes described in this Letter are
barely affected. The actual fabrication of tapered silica fibers
threaded by metallic wires can proceed according to the recently
developed~\cite{sazio_science06}high pressure chemical vapor
deposition technique.

In conclusion, we have demonstrated the possibility of a novel
deeply sub-wavelength multi-channel endoscope based on an array of
metallic wires. The device is based on the remarkable propagation
properties of the transverse electromagnetic (TEM) waves: their
lack of the cutoff and transverse dispersion. Such endoscopes may
find a variety of applications in the areas of infrared imaging,
guiding, and focusing. This work is supported by the ARO MURI
W911NF-04-01-0203 and the AFOSR MURI FA9550-06-1-0279.

%\bibliography{endo_bib}% Produces the bibliography via BibTeX.

\newpage

\newpage

\newpage

\end{document}